# Charge Transport in DNA-based Devices

DANNY PORATH

*Department of Physical Chemistry, Institute of Chemistry, The Hebrew University, Jerusalem 91904, Israel*

GIANAURELIO CUNIBERTI

*Institute for Theoretical Physics, University of Regensburg, 93040 Regensburg, Germany*

ROSA DI FELICE

*INFM Center on nanoStructures and bioSystems at Surfaces (S3), Università di Modena e Reggio Emilia, Via Campi 213/A, 41100 Modena, Italy*

## Table of Contents

# Abstract

Charge migration along DNA molecules has attracted scientific interest for over half a century. Reports on possible high rates of charge transfer between donor and acceptor through the DNA, obtained in the last decade from solution chemistry experiments on large numbers of molecules, triggered a series of direct electrical transport measurements through DNA single molecules, bundles and networks. These measurements are reviewed and presented here. From these experiments we conclude that electrical transport is feasible in short DNA molecules, in bundles and networks, but blocked in long single molecules that are attached to surfaces. The experimental background is complemented by an account of the theoretical/computational schemes that are applied to study the electronic and transport properties of DNA-based nanowires. Examples of selected applications are given, to show the capabilities and limits of current theoretical approaches to accurately describe the wires, interpret the transport measurements, and predict suitable strategies to enhance the conductivity of DNA nanostructures.

# List of Abbreviations

| | |
|---|---|
| Ade (A) | Adenine |
| Cyt (C) | Cytosine |
| Gua (G) | Guanine |
| Thy (T) | Thymine |
| 1D | one-dimensional |
| AFM | Atomic Force Microscope |
| BLYP | Becke-Lee-Yang-Parr (GGA) |
| BZ | Brillouin Zone |
| CNT | Carbon Nanotube |
| DFT | Density Functional Theory |
| DOS | Density of States |
| EFM | Electrostatic Force Microscope |
| GGA | Generalized Gradient Approximation |
| HF | Hartree-Fock |
| HOMO | Highest Occupied Molecular Orbital |
| LEEPS | Low-Energy Electron Point Source |
| LUMO | Lowest Unoccupied Molecular Orbital |
| MP2 | Møller-Plesset $2^{nd}$ order |
| NMR | Nuclear Magnetic Resonance |
| PBE | Perdew-Burke-Ernzerhof (GGA) |
| PW91 | Perdew-Wang 1991 (GGA) |
| SEM | Scanning Electron Microscope |
| SFM | Scanning Force Microscope |
| STM | Scanning Tunneling Microscope |
| TB | Tight Binding |
| TEM | Transmission Electron Microscope |

# 1 Introduction

## 1.1 Devices go molecular – the emergence of molecular electronics

The progress of the electronic industry in the past few decades was based on the delivery of smaller and smaller devices and denser integrated circuits, which ensured the attainment of more and more powerful computers. However, such a fast growth is compromised by the intrinsic limitations of the conventional technology. Electronic circuits are currently fabricated with complementary-metal-oxide-semiconductor (CMOS) transistors. Higher transistor density on a single chip means faster circuit performance. The trend towards higher integration is restricted by the limitations of the current lithography technologies, by heat dissipation and by capacitive coupling between different components. Moreover, the down-scaling of individual devices to the nanometer range collides with fundamental physical laws. In fact, in conventional silicon-based electronic devices the information is carried by mobile electrons within a band of allowed energies according to the semiconductor bandstructure. However, when the dimensions shrink to the nanometer scale, and bands turn into discrete energy levels, then quantum correlation effects induce localization.

In order to pursue the miniaturization of integrated circuits further [1], a novel technology, which would exploit the pure quantum mechanical effects that rule at the nanometer scale, is therefore demanded. The search for efficient molecular devices, that would be able to perform operations currently done by silicon transistors, is pursued within this framework. The basic idea of molecular electronics is to use individual molecules as wires, switches, rectifiers and memories [2-6]. Another conceptual idea that is advanced by molecular electronics is the switch from a top-bottom approach, where the devices are extracted from a single large-scale building block, to a bottom-up approach in which the whole system is composed of small basic building blocks with recognition, structuring and self-assembly properties. The great advantage of molecular electronics in the frame of the continued device miniaturization is the intrinsic nanoscale size of the molecular building blocks that are used in the bottom-up approach, as well as the fact that they may be synthesized in parallel in huge quantities and at low cost. Different candidates for molecular devices are currently the subject of highly interdisciplinary investigation efforts, including small organic polymers [6-11], large bio-molecules [12-20], nanotubes and fullerenes [21-24]. In the following, we focus on the exploration of DNA bio-molecules as prospective candidates for molecular electronic devices.

For the scientists devoted to the investigation of charge mobility in DNA, a no less important motivation than the strong technological drive is that DNA molecules comprise an excellent model system for charge transport in one-dimensional polymers. This most well-known polymer enables an endless number of structural manipulations in which charge transport mechanisms like hopping and tunneling may be studied in a controlled way.

## 1.2 The Unique Advantages of DNA-Based Devices – Recognition and Structuring

Two of the most unique and appealing properties of DNA for molecular electronics are its double-strand *recognition* and a special structuring that suggests its use for *self-assembly*.

Molecular *recognition* describes the capability of a molecule to form selective bonds with other molecules or with substrates, based on the information stored in the structural features of the interacting partners. Molecular recognition processes may play a key role in molecular devices by: (a) driving the fabrication of devices and integrated circuits from elementary building blocks, (b) incorporating them into supramolecular arrays, (c) allowing for selective operations on given species potentially acting as dopants, and (d) controlling the response to external perturbations represented by interacting partners or applied fields.

*Self-assembly*, which is the capability of molecules to spontaneously organize themselves in supramolecular aggregates under suitable experimental conditions [25], may drive the design of well structured systems. Self-organization may occur both in solution and in the solid state, through hydrogen-bonding, Van der Waals and dipolar interactions, and by metal-ion coordination between the components. The concept of selectivity, on which both recognition and self-assembly are based, originates from the concept of information: that is, the capability of selecting among specific configurations reflects the information stored in the structure at the molecular level. It is natural and appealing to use such features to design molecular devices capable of processing information and signals.

By virtue of their recognition and self-assembling properties, DNA molecules seem particulary suitable as the active components for nano-scale electronic devices [26-28]. DNA's natural function of information storage and transmission, through the pairing and stacking characteristics of its constituent bases, stimulates the idea that it can also carry an electrical signal. However, despite the promising development that has been recently achieved in controlling the self-assembly of DNA [29-32] and in coupling molecules to metal contacts [12,33], there is still a great controversy around the understanding of its electrical behavior and of the mechanisms that might control charge mobility through its structure [34].

The idea that double-stranded DNA, the carrier of genetic information in most living organisms, may function as a conduit for fast electron transport along the axis of its base-pair stack, was first advanced in 1962 [35]. Instead, later low-temperature experiments indicated that radiation-induced conductivity can only be due to highly mobile charge carriers migrating within the frozen water layer surrounding the helix, rather than through the base-pair core [36]. The long lasting interest of the radiation community [37] in the problem of charge migration in DNA was due to its relevance for the mechanisms of DNA oxidative damage, whose main target is the guanine (Gua) base [38]. Recently, the interest in DNA charge mobility has been revived and extended to other interdisciplinary research communities. In particular, the issue of electron and hole migration in DNA has become a hot topic [39,40] for a number of chemistry scholars following the reports that photoinduced electron transfer occurred with very high and almost distance-independent rates between donor and acceptor intercalators along a DNA helix [41,42]. This evidence suggested that double-stranded DNA may exhibit a "wirelike" behavior [43].

From the large body of experimental studies performed in solution that became available in the last decade and appeared in recent reviews [44,45], several mechanisms were proposed for DNA-mediated charge migration, depending on the energetics of the base sequence and on the overall structural aspects of the system under investigation. These mechanisms include single-step superexchange [41], multistep hole hopping [46], phonon-assisted polaron hopping [47], and polaron drift [48]. The above advances drove the interest in DNA molecules also for nanoelectronics. In this field, by virtue of their sequence-specific recognition properties and related self-assembling capabilities, they might be employed to wire the electronic materials in a programmable way [12,13]. This research path led to a set of direct electrical transport measurements. In the first reported measurement, μm-scale λ-DNA molecules were found to be "practically insulating" [12]. However, the possibility that double-stranded DNA may function as a one-dimensional conductor for molecular electronic devices has been rekindled by other experiments, where, e.g., anisotropic conductivity was found in an aligned DNA cast film [49], and ohmic behavior with high conductivity was found also in a 600-nm-long λ-DNA rope [50].

The above measurements, complemented by other experiments which are discussed in section 2, highlight that, despite the outstanding results that have been recently achieved in controlling the self-assembly of DNA onto inorganic substrates and electrodes, there is currently no unanimous understanding of its electrical behavior and of the mechanisms that might control charge mobility through its structure. Our purpose in this chapter is to review the main experiments that have been performed to measure directly the conductivity of DNA molecules, and to correlate the measurements to the state-of-the-art theoretical understanding of the fundamental electronic and transport features.

## 1.3 Charge transport in device configuration versus charge transfer in solution chemistry experiments

As already outlined, the interest in the charge migration through DNA grew in three different scientific communities in an almost historical path. The problem originated from the study of genetic mutations related to cancer therapy [37,38]. It was then re-framed in the spirit of determining how fast and how far can charge carriers migrate along the DNA helix in solution [43]. Finally, it was re-formulated again in the nanoscience field to question whether such charge motions are capable of inducing large enough currents in DNA-based electronic devices in a dry environment (namely, with the molecules in conditions very different from the native biochemical ones). These research lines proceed separately but bear connections that may finally unravel a uniform vision and interpretation for the mechanisms that control the motion of charge carriers in various DNA molecules. However, care should be taken in advancing a unique paradigm for the interpretation of data coming from different investigation schemes.

Here, we aim at elucidating how the problem is formulated within the "solution chemistry" community and the "solid state" community, and how the experimental investigations are conducted. The theories related to the different classes of measurements are mentioned later in section 3. The theoretical foundations of the relationship between physical observables revealed in solution and in the electrical transport experiments have been recently thoroughly formulated by Nitzan in different regimes for charge mobility (one-step superexchange – tunneling – and multi-step hopping) [51,52].

The experiments in solution, based on electrochemistry techniques, are targeted at measuring electron-transfer rates between a donor and an acceptor as a function of the donor-acceptor distance and of the interposed base sequence. The donor is a site along the base stack, where a charge (usually positive, forming a radical cation or "hole") is purposely injected into the structure, and the acceptor is a "hole trapping" site at a given distance. The results are an average signal measured over a large number of molecules. The interpretation is generally given in terms of the change of localization site for the hole. The inherent structure of the molecule is compromised by the transfer process, in the sense that the charge state at distinct sites along the helix before and after the hole migration is different. In these experiments there is no tunneling barrier for the charge to overcome when injected into the molecule.

The experiments in the solid state are based on several techniques, including imaging, spectroscopy, and electrical transport measurements that reveal the electric current flux through the molecule under an external field. The results pertain to single molecules (or bundles) and can be re-measured many times. The roles of the donor and of the acceptor are in this case played either by the metal leads, or by the substrate and an imaging metal tip. The interpretation is generally given in terms of *conductivity*, determined by the electronic energy levels (if the molecular structure supports the existence of localized orbitals and *discrete* energy levels) or band-structure (if the intramolecular interactions support the formation of delocalized states described by *continuous* energy levels, i.e., dispersive bands). The donor and the acceptor are reservoirs of charges and this fact allows to leave the charge state along the helix unaltered. It is not specified *a-priori* if the mobile charges are electrons or holes: this depends on the availability of electron states, on their filling, and on the alignment to the Fermi levels of the reservoirs.

In both the indirect electrochemical and the direct transport measurements, the electronic structure of the investigated molecules is important [51,52]. It determines the occurrence of direct donor-acceptor tunneling or of thermal hopping of elementary charges or polarons. Direct tunneling can occur either "through-space" if the DNA energy levels are not aligned with the initial and final charge sites or reservoirs, or "through-bond" if they are aligned and modulate the height and width of the tunneling barrier. In the case of tunneling, the bridging bases do not offer intermediate residence sites for the moving charges. On the contrary, in the case of thermal coupling and hopping, the moving charges physically reside for a finite relaxation time in intermediate sites at base planes between the donor and the acceptor along their path, although this may cost structural reorganization energy. Whether the inherent DNA electronic structure is constituted of dispersive bands or of discrete levels may be revealed only in the solid-state experiments. In fact, for the motion of individual charges injected into free molecules in solution, probed by electrochemistry tools, it is not important whether such charges find in the molecules a continuum of energy levels or discrete levels available to modulate the tunneling barrier. This is because only the modulation of the tunneling barrier or the donor-bridge-acceptor coupling can be detected. Alternatively, in direct electrical transport measurements, where charges are available in reservoirs (the metal electrodes), it makes a difference if there is a continuum of electron states or discrete levels in the molecular bridge that are available for mobile carriers. For the ideal case of ohmic contacts, a continuum in the molecule will be manifested in smoothly rising current-voltage curves, whereas for discrete levels the measured *I-V* curves will be step-like revealing quantization.

This chapter is devoted to a review of the latter class of experiments, complemented by the analysis of the theoretical interpretation of the measurements and underlying phenomena.

## 2 Direct electrical transport measurements in DNA

A series of direct electrical transport measurements *through* DNA molecules that commenced in 1998 was motivated by new technological achievements in the field of electron beam lithography and scanning probe microscopy, as well as by encouraging experimental data suggesting high electron-transfer rates. The latter were based on the interpretation of results of charge-transfer experiments conducted on large numbers of very short DNA molecules in solution, in particular by Barton's group at Caltech and by other colleagues [39-47,53-58]. In a perspective it seems now that care should be taken when projecting from those experiments on the electron transport properties of various single DNA molecules in different situations and structures, e.g., long vs short, on surfaces vs suspended, in bundles vs single, in various environmental conditions like dry environment, or in other exotic configurations.

Few works have been published since 1998 describing direct electrical transport measurements conducted on single DNA molecules [12,14,33,50,59-63]. In such measurements one has to bring (at least) two metal electrodes to a physical contact with a single molecule, apply voltage and measure current (or vice versa). Poor intrinsic conductivity, which seems to be the case for DNA, provides a small measured signal. In such cases the electrode separation should be small, preferably in the range of few to tens of nanometers, yet beyond direct tunneling distance and without any parallel conduction path. The performance of these experiments is highly sophisticated and therefore it is not surprising that the number of the reported investigations is small. Performing good and reliable experiments on single segmented molecules is extremely hard but their interpretation on the basis of the current data is even harder. Not only that - each segmented molecule – a polymer – is intrinsically different from the others in the specific details of its structure. Therefore, also the details of its properties bear some uniqueness. Moreover, the properties of these molecules are sensitive to the environment and environmental conditions, e.g. humidity, buffer composition etc. Another difficulty that arises in these measurements is that the contacts to a single molecule, as to any other small system, are very important for the transport but hard to perform and nearly impossible to control microscopically. For example, the electrical-coupling strength between the molecule and the electrodes will determine whether a Coulomb blockade effect (weak coupling) or a mixing of energy states between the molecule and the electrodes (strong coupling) is measured. In the case of weak coupling, the size and chemical nature of the molecule between the electrodes will determine the relative contributions of Coulomb blockade phenomena and of the intrinsic energy gap of the molecule to the current-voltage spectra. For the outlined reasons, we find a large variety in the results of the few reported experiments, most of which done by excellent scientists in leading laboratories.

The question whether DNA is an insulator, a semiconductor or a metal is often raised. This terminology originates from the field of solid-state physics where it refers to the electronic structure of semi-infinite periodic lattices. It is even successfully used to describe the electrical behavior of one-dimensional wires like carbon nanotubes, where a coherent bandstructure is formed. However, it is questionable whether or not this notion describes well, with a similar meaning, the orbital-energetics and the electronic transport through one-dimensional soft polymers that are formed of a large number of sequential segments. In these polymers the number of junctions and phase-coherent "islands" is large and may determine the electronic structure and the transport mechanisms along the wire. In some cases it may be those junctions that constitute a bottleneck for the transport. They will determine the overall electric response of the polymer, in spite of suitable energy levels and/or "bands" in the islands that connect those junctions, that could otherwise enable a coherent charge transport. In the case of a strong coupling between the islands along the polymer, a complex combination of the molecular electron states and of the coupling strengths at the junctions will determine the electrical response of the wire.

DNA in particular is sometimes said to be an insulator or a semiconductor. If we assume the possible formation of a long phase-coherent portion, then it may be useful to introduce a distinction between the two terms. In the bulk the difference between a wide-bandgap semiconductor and an insulator is mainly quantitative with regard to the resistivity. For DNA and other polymers we may instead introduce the following distinction. If we apply a voltage (even high) across a wide-bandgap polymer and successfully induce charge transport through it *without* changing the polymer structure and its properties *in an irreversible way* then it would be a wide-bandgap semiconductor. However, if the structure is *permanently damaged* or changed upon this voltage application then it is an insulator. This distinction is important with regard to the relevant experiments, where very high fields are present, and the methods to check whether the conduction properties of the molecule are reproducible.

In sections 2.1 and 2.2 we will review the direct electrical transport experiments reported on DNA single molecules, bundles, and networks.

## 2.1 Single molecules

The first direct electrical transport measurement on a single, 16-μm-long λ-DNA, was published in 1998 by Braun et al. [12]. In this fascinating experiment the λ-DNA was stretched on a mica surface and connected to two metal electrodes, 12 μm apart. This was accomplished using the double-strand recognition between a short single-strand (hang-over) in the end of the long λ-DNA and a complementary single-strand that was connected to the metal electrode on each side of the molecule (see Fig. 1). Electrical transport measurements through the single molecule that was placed on the surface yielded no observable current up to 10 V.

Later on in 1999 Fink et al. [50] reported ohmic behavior in λ-DNA molecules with a resistance in the MΩ range. The molecules were a few hundred nanometers long and were stretched across ~2 μm wide holes in a metal-covered transmission electron microscope (TEM) grid, as shown in Fig. 2. This fantastic technical accomplishment was achieved in a high-vacuum chamber where a holographic image was created with a low-energy electron point source (LEEPS) claimed not to radiatively damage the DNA. Note, however, that the bright parts of the DNA in the images may suggest scattering of the beam electrons from the molecule, which may indicate the presence of scattering points along the DNA that could effect the charge transport along the molecule. The actual measurement was performed between a sharp tungsten tip, which was connected to the stretched molecule in the middle of one of the grid holes, and the metal covering the TEM grid. The tungsten tip was aligned using the holographic image. An ohmic behavior was observed in the current-voltage (*I-V*) curves, sustained up to 40 mV and then disappeared. The resistance division between two DNA branches appeared consistent with the ohmic behavior. This result seemed very promising. However, while conduction over long distances was observed later in bundles, it was not repeated in further measurements of single DNA molecules with one exception of a superconducting behavior that is discussed later [61]. The resolution of the LEEPS in this measurement did not enable to determine whether it was a single molecule or a bundle that was suspended between the metal tip and the metal grid.

In a further experiment published in 2000 by Porath et al. [14], electrical transport was measured through 10.4-nm-long (30 base-pairs) homogeneous poly(dG)-poly(dC) molecules that were electrostatically trapped [64-65] between two Pt electrodes (see Fig. 3). The measurements were performed at temperatures ranging from room temperature down to 4 K. Current was observed beyond a threshold voltage of 0.5-1 V suggesting that the molecules transported charge carriers. At room temperature in ambient atmosphere, the general shape of the current-voltage curves was preserved for tens of samples but the details of the curves varied from curve to curve. The possibility of ionic conduction was ruled out by measurements that were performed in vacuum and at low temperature, where no ionic conduction is possible. High reproducibility of the *I-V* curves was obtained at low temperature for tens of measurements on a certain sample, followed by a sudden switching to a different curve-shape (see inset of Fig. 4) that was again reproducible (e.g., peak position and height in the *dI/dV* curves, Fig. 4). This variation of the curves in different samples can originate from the individual structural conformation of each single molecule, or from the different formation of the specific contact. The variation of the curves measured on the same sample may be also due to switching of the exact overlap of the wavefunctions that are localized on the bases. A rather comprehensive set of control experiments helped to verify the results and ensure their validity. The existence of the DNA between the electrodes was verified by incubating the DNA devices with DNase I, an enzyme that specifically cuts DNA (and not any other organic or inorganic material). Following incubation of the sample with the enzyme the electrical signal was suppressed, indicating that the molecule through which the current was measured before is indeed DNA. The procedure was cross-checked by repeating this control experiment in the absence of Mg ions in the enzyme solution so that the action of the enzyme could not be activated. In this case the signal was not affected by the incubation with the enzyme. This procedure ensured that it was indeed the enzyme that did the cut (see Fig. 3b), thus confirming again that it was the DNA between the electrodes. This experiment clearly proves that at least short homogeneous DNA molecules are capable of transporting charge carriers over a length of about 10 nm.

Additional experiments were performed in 2001 in the same laboratory by Storm et al. [60], in which longer DNA molecules (> 40 nm) with various lengths and sequence compositions were stretched on the surface between planar electrodes in various configurations (see Fig. 5). No current was observed in these experiments suggesting that charge transport through DNA molecules longer than 40 nm on surfaces is blocked.

In parallel, Kasumov et al. [61] reported ohmic behavior of the resistance of λ-DNA molecules deposited on a mica surface and stretched between rhenium-carbon electrodes (see Fig. 6). This behavior was measured at temperatures ranging from room temperature down to 1 K. Below 1 K a particularly unexpected result was observed: proximity-induced superconductivity. The resistance was measured directly with a lock-in technique and no current-voltage curves were presented. This surprising proximity-induced superconductivity is in contrast to all the other data published so far, and with theory. No similar result was reported later by this or any other group.

In another attempt to resolve the puzzle around the DNA conduction properties, de Pablo et al. [59] applied a different technique to measure single λ-DNA molecules on the surface in ambient conditions. They deposited many DNA molecules on mica, covered some of them partly with gold, and contacted the other end of one of the molecules (>70 nm from the electrode) with a metal AFM tip (see Fig. 7). No current was observed in this measurement. Furthermore, they covered ~1000 parallel molecules on both ends with metal electrodes and again no current was observed. Yet another negative result published in 2002 was obtained in a similar experiment by Zhang et al. [33] who stretched many single DNA molecules in parallel between metal electrodes and measured no current upon voltage application. Both results were consistent with the Storm experiment [60]. Beautiful and quite detailed measurements with different results on shorter molecules were reported by Watanabe et al. [62] and Shigematsu et al. [63] using a rather sophisticated technique. A short, single DNA molecule was contacted with a triple-probe AFM. The DNA molecule was laid on the surface and contacted with a triple-probe AFM consisting of 3 conducting CNTs. Two of them, 20 nm apart, were attached to the AFM (see Fig. 8a). In one case, voltage was applied between the nanotube on one side of the molecule and the tip-nanotube that contacted the DNA molecule at a certain distance from the side electrode, so that the dependence of the current on the DNA length was measured under a bias voltage of 2 V between the two electrodes. The current dropped from 2 nA at ~2 nm to less than 0.1 nA in the length range of 6 to 20 nm. In the second experiment reported by this group [63], current was measured between the side nanotubes (20 nm apart) under a bias voltage of 2 V upon moving the tip-nanotube that served this time as a gate along the DNA molecule. A clear variation of the current due to the effect of the gate electrode, reproducible forwards and backwards, is observed. The current-voltage curves in this experiment are measured through carbon nanotubes. Their conductivity is indeed much higher than that of the DNA molecule and therefore likely to have only a small effect on the *I-V*'s. However, this and the contacts of the nanotubes to the AFM tip and metal electrodes still might have an effect on the measured results.

From the direct electrical transport measurements on single DNA molecules reported so far one can draw some very interesting conclusions. First, it is possible to transport charge carriers through single DNA molecules, both homogeneous and heterogeneous. This was observed, however, only for short molecules in the range of up to 20 nm in the experiments of Porath et al. [14], Watanabe et al. [62] and Shigematsu et al. [63]. All the three experiments demonstrated currents of order 1 nA upon application of voltage of ~1 V. The experiments by Fink et al. [50] and Kasumov et al. [61] showed higher currents and lower resistivities over longer molecules (hundreds of nm), but they were never reconfirmed for individual molecules. In all the other experiments by Braun et al. [12], de Pablo et al. [59], Storm et al. [60] and Zhang et al. [33] that were conducted on long (>40 nm) single DNA molecules attached to surfaces no current was measured. This result is not too surprising if we recall that DNA is a soft segmented molecule and is therefore likely to have distortions and defects when subjected to the surface force field. This is also manifested in AFM imaging where the measured height of the molecule is different from its "nominal height" [60,66], partly due to the effect of the pushing tip and partly due to the effect of the surface force field. This force field may be a reason for blocking the current but not necessarily the only one.

The conclusion of poor conductivity in long single molecules on surfaces is further supported by indirect electrostatic force microscope (EFM) measurements, reported by Bockrath et al. [66] and Gómez-Navarro et. al. [67]. In these measurements no attraction was found between a voltage-biased metal-tip and the λ-DNA molecules lying on the surface. This indicates that the electric field at the tip failed to induce long-range polarization in the molecules on the surface, which would in turn indicate charge mobility along the molecule, as was found for carbon nanotubes.

## 2.2 Bundles and Networks

A few measurements of direct electrical transport were performed also on single bundles. Other measurements were done on networks formed of either double-stranded DNA [68] or alternative poly-nucleotides [69]. All the reported measurements showed current flowing through the bundles. We will show a few examples here.

The most productive group in the "networks field" is the group of Tomoji Kawai from Osaka that published an extended series of experiments on different networks and with various doping methods [70,71, and references therein]. In one of their early experiments they measured the conductivity of a single bundle [68]. This was done in a similar way to the de Pablo experiment [59] (see Fig. 7), placing the bundle between a metal-covered AFM tip on one side of the molecule and under a metal electrode that covered the rest of the bundle (see Fig. 9). The conductivity of a poly(dG)-poly(dC) bundle was measured as a function of length (50-250 nm) and was compared with that of a poly(dA)-poly(dT) bundle. The results showed a very clear length-dependent conductivity that was about an order of magnitude larger for the poly(dG)-poly(dC) bundle.

One of the interesting measurements among the "bundle experiments" was done by Rakitin et al. [72]. They compared the conductivity of a λ-DNA bundle to that of an M-DNA [72-75] bundle (DNA that contains an additional metal ion in each base-pair, developed by the group of Jeremy Lee from Saskatchewan). The actual measurement was performed over a physical gap between two metal electrodes in vacuum (see Fig. 10). Metallic-like behavior was observed for the M-DNA bundle over 15 μm, while for the λ-DNA bundle a gap of ~0.5 V in the *I-V* curve was observed followed by a rise of the current.

Another measurement that follows the line of the Porath et al. [14] experiment was performed by Yoo et al. [76]. In this experiment, long poly(dG)-poly(dC) and poly(dA)-poly(dT) molecules were electrostatically trapped between two planar metal electrodes that were 20 nm apart (see Fig. 11) on a $SiO_2$ surface, such that they formed a bundle that was ~10 nm wide. A planar gate electrode added another dimension to this measurement. The current–voltage curves showed a clear current flow through the bundle and both temperature and gate dependencies. The resistivity for the poly(dG)-poly(dC) was calculated to be 0.025 Ωcm.

An interesting experiment on a DNA-based network embedded in a cast film had already been done by Okahata et al. already in 1998 [49]. In this pioneering experiment the DNA molecules were embedded (with side groups) in a polymer matrix that was stretched between electrodes (see Fig. 12). It was found that the conductivity parallel to the stretching direction (along the DNA) was ~4.5 orders of magnitude larger than the perpendicular conductivity.

In a recent experiment that was mentioned above with regard to single molecules measurements, Shigematsu et al. [63] prepared a more complex network that included acceptor molecules. They found a network conductivity that increased with the guanine content.

Measurements on a different type of DNA-based material were reported by Rinaldi et al. [69] (see Fig 13). In this experiment they deposited a few layers of deoxyguanosine ribbons in the gap between two planar metal electrodes, ~100 nm apart. The current-voltage curves showed a gap followed by rise of the current beyond a threshold of a few volts. The curves depended strongly on the concentration of the deoxyguanosine in the solution.

## 2.3 Conclusions from the experiments about DNA conductivity

More and more evidences accumulating from the direct electrical transport measurements show that it is possible to transport charge carriers along short single DNA molecules, in bundles of molecules and in networks, although the conductivity is rather poor. This is consistent with the picture that emerges from the electron-transfer experiments. By this picture, that is becoming a consensus, the two most fundamental electron-transfer processes are coherent tunneling over a few base-pairs and diffusive thermal hopping over a few nanometers. However, transport through long single DNA molecules (>40 nm) that are attached to the surface is apparently blocked. It may be due to the surface force field that induces many defects in the molecules and blocks the current or any additional reason.

Therefore, if one indeed wants to use DNA as an electrical molecular wire in nanodevices, or as a model system for studying electrical transport in a single one-dimensional molecular wire, then there are a few possible options. One option is to use doping by one of the methods that are described in the literature [71-75] (e.g., addition of intercalators, metal ions or $O_2$ etc.). Another way is to reduce the surface affinity of the DNA molecules and hence the effect of the surface force field (e.g., by a pre-designed surface layer) on the attached DNA. Yet another way could be to use more exotic structures such as DNA quadruple-helices instead of the double-stranded structure. Such constructions may offer an improved stiffness and electronic overlap that may enhance the conductivity of these molecules.

# 3 Theoretical understanding of charge transport in DNA-based wires

The theoretical approaches that were applied so far to the study of charge mobility in DNA molecules can be divided into two broad classes.

(i) The kinetic determination of the charge-transfer rates between specific locations on the base sequence, after the Marcus-Hush-Jortner theory [77,78], is the preferred route by the (bio)chemistry community. In these approaches, the electronic structure information is employed only at the level of individual bases or couples of stacked neighboring bases. The results obtained may be compared to the measured charge-transfer rates, and employed to devise models for the mobility mechanisms, addressing dynamical processes by which the charges might move along the helix, e.g., one-step superexchange, hopping, multiple hopping, polaron hopping [41,46-48,79-81].

(ii) The computation of the molecular electronic structure for model and real extended DNA-base aggregates, which affects the quantum conductance and hence the quantities directly measured in transport experiments, is instead linked to the investigations performed by the nanoscience community to explore the role and the efficiency of DNA in electronic devices. The results of such calculations may help devise models for charge mobility from a different point of view, e.g., to unravel the role of the electronic structure in determining the shape of the measured current-voltage characteristics.

The two approaches are not unrelated and a complementary analysis of both kinds of studies would finally shed light onto the detailed mechanisms for charge migration along DNA wires [51,52]. The kinetic theories are reviewed in other chapters of this book. Here, we focus on results obtained for the electronic structure of extended DNA-base stacks, and describe their influence on the electrical conductivity of DNA-based nanostructures.

## 3.1 Methods to study quantum transport at the molecular scale

In principle, one would like to perform accurate computations of the relevant measurable quantities to assess the conductivity of the fabricated molecular devices. For coherent transport in the absence of dissipative scattering, the Landauer theory [82-84] is a well defined frame. It allows to describe the quantum conductance and the current-voltage characteristics of the system in the device configuration between metallic leads, when the quantum electron structure of the system molecule+leads is known. However, the most manageable formulation of the theory, based on the computation of the Green function (electron propagator), does not allow a straightforward interplay with first-principle methods that are applied to calculate the molecular electronic structure (except for very recent formulations [85-88] that are still very cumbersome and have not yet been applied to DNA-based wires). Therefore, we split our review of the theoretical investigations in two sets. One set is devoted to the parameter-free determination of the electronic structure, without the extension to the measurable quantities (discussed in section 3.2). A second set devoted to the mesoscopic measurable quantities (such as the *I-V* curves) with the input electronic structure based on empirical calculations (section 3.3). The former allows a thorough understanding of the basic physico-chemical mechanisms, whereas the latter allows for a direct comparison with the device experiments.

<u>Electronic structure from first principles.</u> Among the different possible methods to study the electrical conductivity of solid state devices, the deepest insight into the process might be gained by studying the energy levels and wavefunctions (or, alternatively for bulk materials, the bandstructure).

The most sophisticated quantum chemistry computational techniques that have been applied to nucleotides are based on the determination of the structure at the Hartree-Fock (HF) level, which exactly includes Coulomb exchange effects, but totally neglects correlations. Correlation effects are then taken into account with the application of second-order Møller-Plesset perturbation theory (MP2) to compute relative formation energies [89,90] with a high degree of accuracy. These cumbersome studies, conventionally named as MP2//HF, provide an accurate determination of the geometry and energetics of stacked and hydrogen-bonded base pairs, but do not presently allow the extension to more complex aggregates. Real nucleotide structures are not accessible to them and require more drastic approximations.

One interesting scheme based on Density Functional Theory (DFT) is particularly appealing, because with the current power of the available computational facilities it enables the study of reasonably extended systems. DFT has been applied with a variety of basis sets (atomic orbitals or plane-waves) and potential formulations (all-electron or pseudopotentials) to complex nucleobase assemblies, including model systems [91-93] and realistic structures [59,94-96]. DFT [97-99] is in principle an ab-initio approach, as well as MP2//HF. However, its implementation in manageable software requires some approximations. The most drastic of all the approximations concerns the exchange-correlation (xc) contribution to the total DFT functional, which is described in a mean-field approach. Whereas the first widely used Local Density Approximation (LDA) functional performs extremely well in bulk contexts, it is not able of quantitatively describing reactive chemistry. Improved Generalized-Gradient-Approximation (GGA) [100-102] and hybrid [103-105] functionals are now able to provide an excellent description of the properties of many-electron systems in molecular environments.

Quantum transport. In order to obtain estimates of quantum transport at the molecular scale [106], electronic structure calculations must be plugged into a formalism which would eventually lead to observables such as the linear conductance (equilibrium transport) or the current-voltage characteristics (non-equilibrium transport). The directly measurable transport quantities in mesoscopic (and *a fortiori* molecular) systems, such as the linear conductance, are characterized by a predominance of quantum effects — e.g. phase coherence and confinement in the measured sample. This was firstly realized by Landauer [82] for a so-called "two terminal" configuration where the sample is sandwiched between two metallic electrodes energetically biased to have a measurable current. Landauer's great intuition was to relate the conductance to an elastic scattering problem and thus to quantum transmission probabilities. Most implementations of conductance calculations were so far developed for describing phase coherent systems, typically semiconductor heterostructures. The latter are fabricated at the micron/submicron scale, a size large enough to justify an approximate treatment of the electronic structure, typically operated by employing a tight-binding (TB) Hamiltonian. However, even with certain classes of smaller and truly molecular systems, an empirical TB treatment of the electronic structure already provides excellent qualitative and in some cases quantitative predictions. This is the case of carbon nanotubes (CNTs) where a simple TB Hamiltonian (including a single $\pi$-orbital per carbon atom) is enough to classify a metallic or semiconducting behavior depending on the CNT chirality [107]. In some cases, as in complex structures like DNA wires, the choice of embracing an approximated electronic structure is definitely convenient in order to obtain analytical treatments which might guide the understanding of the basic physics of the system, as Section 3.3 presents for the experiment by Porath et al. [14].

## 3.2 Electronic structure of nucleobase-assemblies by first principles

After briefly presenting some important milestones of MP2//HF studies in the quantum chemistry description of DNA base pairs, we turn to a more extensive discussion of DFT results for extended DNA-base aggregates, including model stacks and real molecular fragments.

Quantum Chemistry. The calculations performed by Šponer and coworkers to evaluate the structure and stability of hydrogen-bonded [89] and stacked [90] base pairs should be retained as important milestones for the application of first-principle computational methods to interacting nucleotides. The authors demonstrated that this kind of theoretical analysis is able to reproduce many of the experimental features, and has a predictive power. Moreover, the description of hydrogen-bonded complexes [89] of DNA bases is propaedeutic to the development of any empirical potential to model DNA molecules and their interaction with drugs and proteins. The conclusions of their investigations may be summarized in the following information: (i) structure of the most favorable hydrogen-bonded and stacked dimers; (ii) rotation- and distance-dependence of the relative energy of stacked pairs; (iii) description of the relevant interactions that determine the relative stability of base-pairs. Concerning the latter issue, it was found that the energetics of stacked pairs is essentially determined by correlation effects, and therefore can only be accessed through a purely quantum chemical description. On the other hand, the energetics of hydrogen-bonded pairs is well described already at the HF level, and also the DFT treatment is reliable in this context (Van der Waals interactions may be added a-posteriori [108]). The investigations by Šponer and coworkers remained limited to the analysis of the structure and energetics of DNA-base pairs. The electronic properties were addressed by quantum chemistry methods mainly at the HF level [109-111] (which completely lacks correlation terms): these calculations are discussed in this book in the chapter by Rösch and Voityuk. The notable exception to this restriction is the work performed by Ladik and coworkers [112,113], who evaluated the shifts of the electron levels and gaps due to correlation effects in the MP2 scheme.

Density Functional Theory. The DFT scheme is more suitable to compute the electronic properties of the extended DNA molecules that are proposed as candidates for electrical wires, and has been successfully applied to a number of different structures. Provided DFT reproduces the main structural features (e.g. bond lengths and angles, stacking distances) in agreement with the MP2//HF calculations, the electronic properties thus derived are fully reliable. DFT is the method of choice for studying reactive chemistry [114].

DFT simulations of DNA-like structures constituted of more than two bases, with an accent on their electronic properties, have become available only recently [59,91-96,115]. In our review, we mainly focus on periodic systems obtained by replicating a given elementary unit. For such periodic assemblies, prototypes of DNA-based wires, it is possible to define a "crystal" lattice in one dimension: This allows the extension of the concept of band dispersion and Bloch-like conduction to molecular wires to which one may assign a periodicity length. The advantage of defining a bandstructure for a DNA-based wire is the possible interpretation of experimental results in terms of conventional semiconductor-based device conductivity, using the concept of a band of allowed energy values within which a delocalized electron is mobile. Indeed, we wish to point out that in some recent theoretical studies, the concepts of a bandstructure and dispersive energy bands were ambiguously introduced [116], whereas in principles one could only speak of energy manifolds [59,94,95].

In reporting the bandstructure calculations for DNA-based molecular wires, we first focus on the kind of information that may be extracted from the study of model nucleobase assemblies, and then analyze the attempts to treat realistic molecules. Finally, we give a brief account of the environmental effects, e.g. the presence of water molecules and counterions.

**3.2.1 Model Base Stacks.**

Di Felice and coworkers performed ab-initio calculations of model systems in the frame of plane-wave pseudopotential DFT-LDA(-BLYP) [91]. They considered periodic homo-guanine stacks, motivated by the particular role played by this base both in solution chemistry experiments (lowest ionization potential) [53,54], and in direct conductivity measurements (sequence uniformity, higher stiffness of G-C pairs with respect to A-T pairs) [14]. Their study aimed at understanding the role of various structural features of the base-chains in the establishment of continuous orbital channels through the G aggregates. A particular focus was given to: (i) the role of the relative rotation angle between adjacent bases along the π-stack, and (ii) the relative role of π-stacking and hydrogen-bonding in structures where both kinds of interactions exist. The results allow to draw some general conclusions about the configurations that may be conducive to the formation of delocalized wire-like orbitals. Although the model structures are only partially related to G-rich DNA duplexes [14], they are directly related to supramolecular deoxyguanosine fibers [117] that are also suggested as potential building blocks for molecular nanodevices [17,69].

The π-like nature of the guanine HOMO (see Figure 14a), namely the protrusion out of the molecular plane, suggests that it might easily hybridize with other similar orbitals in a region of space where a relatively large superposition occurs. The degree of overlap depends on the relative positions of the C-C and C-N bonds where the HOMO charge density is mainly localized, determined by the relative rotation angle of the guanines in consecutive planes of the stack. The calculated interplanar distance in stacked Gua dimers with different relative rotation angles is 3.37 Å in perfect agreement with experimental and Hartree-Fock data [90]. This value was used to fix the periodicity length in the extended guanine chains. A schematic diagram of the model G stacks is shown in Figure 14b.

Among the several relative rotation angles that were considered, the configurations most representative for the discussion about a viable band-like conductivity mechanism in guanine π-stacks are illustrated in Figure 15 (insets), along with the computed bandstructure diagrams. The conclusion that can be drawn from the bandstructure analysis of these model guanine strands is that dispersive bands may be induced only by a large spatial π overlap of the HOMO (LUMO) orbitals of adjacent bases in the periodic stack. Such an overlap is maximum for eclipsed guanines (Fig. 15, left), and very small for guanines rotated by 36º (Fig. 15, right) as in B-DNA. These results suggested that a band-like conductivity mechanism occurring via band dispersion and almost free-like mobile carriers (which should be injected through a suitable doping mechanism) is not viable along frozen G-rich stacks. It cannot be excluded that atomic fluctuations locally induce an amount of overlap larger than in frozen B-DNA, with partial interaction and bandstructure formation at least over a typical coherence length. This is possible for a short length, whereas other dynamical mechanisms should be invoked to explain long-range charge migration. As a final remark, we note that the band dispersion found in the model guanine chains described in this sub-section was solely due to the translational symmetry in the infinite wire, not including the helical symmetry. This remark and possible ambiguities about the way different authors report their results will be clarified in the following sub-section.

### 3.2.2 Realistic DNA-based Nanowires

From the above results, it appears very unlikely that even continuous and uniform base sequences form true semiconducting bands (and associated delocalized orbitals along the helix axis). Nevertheless, there might be other intervening mechanisms of orbital mixing, characteristic of supramolecular structures and without an exact equivalent in the solid state inorganic crystals, that may induce the effective behavior of a semiconductor. We illustrate this idea with a few examples that appeared in the literature [59,94,116], concerning homo-Gua [94] and Gua-Cyt [59,116] aggregates that resemble realistic structures and include both π-stacking and hydrogen-bonding interactions. The mechanism suggested for a semiconducting behavior, alternative to pure crystal-like Bloch conductivity, involves manifolds of localized levels: These manifolds are formed as a consequence of the inter-base interactions that do not involve chemical bonding.

DFT simulations of periodic wires show that the weak coupling between the building-blocks (Gua, or Gua-Cyt pairs) contained in the periodicity length split the energy levels of the coupled orbitals, which should be otherwise degenerate in the absence of inter-planar interaction. Such a splitting results in the appearance of a "band" of closely-spaced energy levels. Although each level is non-dispersive, the complete set of similar orbitals (e.g. HOMO) is gathered into a band with a given amplitude, which plays the same role as a dispersive band if the energy splitting between levels in the manifold is small. The formation of energy manifolds was found in both poly(G)-poly(C) [59,116] and G-quartet [94] wires, for both occupied and empty levels, with amplitudes dependent on the particular molecule and on the computational method. We discuss in the following some quantitative features of the "manifold mechanism" for the origin of a semiconducting bandstructure.

de Pablo and coworkers [59] performed linear-scaling pseudopotential numerical-atomic-orbital DFT-PBE [118] calculations of poly(dG)-poly(dC) DNA sequences with periodicity length corresponding to 11 base planes, in dry conditions [115]. The electronic structure was determined for the optimized geometry. The ordered poly(dG)-poly(dC) wire was characterized by filled and empty "bands" around the Fermi level constituted of eleven states, i.e. one state for each base pair. The highest filled band was derived from the HOMO's localized on the G bases and had a bandwidth of 40 meV (effective dispersion for hole conductivity). The lowest empty band was derived from the LUMO's localized on the C bases and had a bandwidth of 250 meV (effective dispersion for electron conductivity), meaning an effective mass in the range typical of wide-band-gap semiconductors. The LUMO-derived band was separated by an energy gap of 2 eV from the HOMO-derived band: this value is affected by the well-known underestimate of DFT-computed energy gaps [97]. Artacho and corworkers described the establishment of a "band" in terms of the helical symmetry [115]. By virtue of this symmetry, the manifold of energy levels of the Bloch orbitals found at the $\Gamma$ point of the 1D BZ is split into equally spaced reciprocal-space points along the helix axis, giving an effective band dispersion. The effective electron orbitals (shown as isosurfaces in the original work [59]) are obtained as linear combinations of the one-particle computed Bloch orbitals. The authors also noticed that the wide band-gap itself does not rule out electrical conduction, if any doping mechanisms capable of injecting free carriers (e.g. defects in the hydrogen atoms or counterions saturating the phosphates) is active in the molecule. In the same work [59,115], it was shown that a defected poly(dG)-poly(dC) sequence exhibits electronic localization over few base pairs, with consequent coherence breaking and exponential decay of the conductance with length.

A similar behavior was found for a non-periodic 20-base-pair-long poly(G)-poly(C) molecule, by means of Valence Effective Hamiltonian (VEH) calculations, whose accuracy is claimed comparable to that of DFT [116]. In this investigation the DNA double helix was modeled by two separate strands, assuming that the hydrogen-bonding in each plane gives a "weak" contribution to the Gua-Cyt interaction. We wish to point out that the hydrogen-bonding contribution to the energetics is not weak, but stronger than the base stacking contribution [89-91]. A more suitable rephrasing of the concept requires the specification that the weakness of H-bonding is limited to its contribution to inter-base orbital hybridization and delocalization [59,91-92]. Additionally, for the finite poly(G)-poly(C) molecule investigated by Hjort and Stafström the HOMO-derived bandwidth was found to be 0.2 eV. This value is much larger than that found for the periodic wire investigated by de Pablo and coworkers [59], and was not clearly discussed in terms of a manifold of localized orbitals, thus giving the ambiguous interpretation of a real dispersive Bloch-like band.

The semiconducting-like behavior induced by split-level effective bands was recently identified for a G-quartet nanowire [94] which is suggested as an improved structure for an electrical molecular wire. Tubular sequences of G tetramers were investigated by means of pseudopotential plane-wave DFT-BLYP calculations of the equilibrium geometry and electronic structure. In the same way as the base-pairs (G-C and A-T) stack on top of each other to form the inner core of double-helix DNA, the tetrames (G4) are the building blocks of a quadruple-helix form of DNA, labeled as G4-DNA. Each tetrameric unit is a planar aggregate of hydrogen-bonded guanines arranged in a square-like configuration (see Figure 16a) with a diameter of 2.3 nm, slightly larger than the 2.1 nm of native DNA. By stacking on top of each other as shown pictorially in Figure 16b, these G4-DNA planes form a periodic columnar phase with a central cavity that easily accommodates metal ions coordinating the carbonyl oxygen atoms. The G4-DNA quadruple helix was simulated by periodically repeated supercells, containing three stacked G4 tetramers, separated by 3.4 Å and rotated by 30º along the stacking direction (Fig. 16b,c). The starting atomic configuration was extracted from the X-ray structure of the *G-quadruplex d(TG4T)* [119].

The electronic bandstructure of the $K^+$-filled G4 quadruple helix is shown in Figure 17 (left), along with the total DOS (right). The special symmetry of the G4 quadruple helix increases the spatial overlap between consecutive planes with respect to a segment of G-rich B-DNA, thus suggesting (after the discussion in sub-section 3.2.1) a possible enhancement of the band-like behavior. However, the results presented in Figure 17 reveal a different situation. It is found that the inter-plane π superposition is not sufficient to induce the formation of delocalized Bloch orbitals and dispersive energy bands. The bandstructure shows in fact that the bands remain flat, typical of supramolecular systems in which the electron states are localized at the individual molecules of the assembly. Nevertheless, another mechanism for delocalization takes place. The plot in Figure 17 (left) identifies the presence of multiplets (or manifolds), each constituted of 12 energy levels. The 12 electron orbitals associated with a multiplet have identical character and are localized on the 12 guanines in the periodic unit cell. The energy levels in a multiplet are separated by an average energy difference of 0.02 eV, smaller than the room temperature energy $K_BT$: therefore, the coupling with the thermal bath may be sufficient to mix the G-localized orbitals and produce effective delocalization. The resulting DOS in Figure 17 (right) shows that the multiplet splitting effectively induces the formation of dispersive energy peaks. Filled and empty bands are separated by a band-gap of 3.5 eV. The most relevant DOS peak for transport properties is π-like. It is derived from the HOMO manifold, and has a bandwidth of 0.3 eV. Despite the valence bands do not form a continuum, they exhibit dispersive peaks in precise energy ranges, which may host conductive channels for electron/hole motion.

The manifold energy splitting is accompanied by the formation of delocalized orbitals as shown by the contour and isosurface plots in Figure 18: a clear mobility channel is identified at the outer border of the G4 column. In the ground state of the system, all the valence bands are filled and the conduction bands are empty, so that mobile carriers are absent. Therefore, efficient doping mechanisms, that may eventually rely on the native structural properties of these G4-DNA wires and on the presence of cations, should be devised in order to exploit them as electrical conductors. Indeed, we note that in the study of the G4-DNA-like wires [94] the K semi-core states were not taken into account, and that they may be able to provide hybridization with the base stack and an intrinsic doping factor. Such developments move along the direction of investigating the electronic modifications introduced in DNA helices by metal cations inserted in the inner core [72-74,120-121].

### 3.2.3 Effects of counterions and solvation shell

Two recent DFT calculations, performed at the upper limits of the computational power available with the most sophisticated parallel computers, for an infinite wire [95] and for a finite four-base-pair molecule [96], have addressed the static and dynamic role of counterions in the determination of the electron energy levels and wavefunctions.

Gervasio and coworkers analyzed a periodic nucleotide structure obtained from the finite molecule d(G*p*C*p*)*6* in the Z-DNA conformation. The crystal structure of this molecule is known and was assumed as the starting configuration for ab-initio molecular dynamics/quenching simulations, with 1194 atoms in the periodically repeated supercell including the sugar-phosphate backbone, water molecules and Na$^+$ counterions [95]. The positions of the latter, not resolved in the crystal structure, were initially assigned on the basis of considerations about the available volume along the backbone and grooves. All the atoms were then relaxed. Besides interesting results about novel arrangements of water clusters surrounding and penetrating the double helix, the authors report the analysis of the electronic structure. The outstanding outcome of this computation is the evidence for Na$^+$-related electron states within the band-gap occurring between Gua- and Cyt-levels (Figure 19), which would pertain to the electronic structure of the base stack alone. As a critical comment to this work, we point out the following considerations: (i) the description in terms of Gua and Cyt manifolds is in agreement with the calculations described in sections 3.2.1 and 3.2.2; (ii) the effect of the static metal cations does not destroy the Gua-Cyt underlying "bandstructure", but only introduces additional empty electron states, not hybridyizing with the Gua and Cyt orbitals, that might be appealing for doping mechanisms. Therefore, the studies that address the electronic structure of the base core stack remain valid as a fundamental point to understand more complex mechanisms that arise by complicating the geometry.

Another interesting account of Na$^+$ counterions was devoted to their dynamical role [96]. Barnett and coworkers performed classical molecular dynamics simulations of a finite B-DNA duplex d(5'-GAGG-3') with an intact sugar-phosphate backbone, including the neutralizing Na$^+$ counterions and a hydration shell. The classical calculations allowed them to sample the Na$^+$ "visitation map" (i.e. the map of the sites explored by the Na$^+$ counterions during the dynamical evolution), from which selected configurations differing for the positions of the cations (populating either backbone sites or helix grooves) were extracted and described by DFT quantum calculations. From their results, the authors identify an "ion-gated transport" mechanism. This mechanism is based on the fact that the hole, described as a total-charge difference between the charged and the neutral system, becomes localized at different core sites depending on the cation positions. Therefore, by migrating along the molecule axis outside the helix, the metal cations drive the hole hopping between G bases and GG traps. Differently from the other studies described in section 3.2, this latter research is not dedicated to the investigation of band-conduction channels through the establishment of delocalized electron states, but addresses the issue of hole localization at various sites induced by structural fluctuations. We believe that both points of view are of relevance for dc conductivity measurements in the solid state, where in principle ionic motions are more frozen than in solution. To which extent this is true has not yet been established and likely depends on the experimental settings.

Finally, we note that the investigations described in this sub-section address the role of cations external to the DNA helix. To our knowledge, theoretical studies of cations inside the helix, possibly modifying the base pairing through electronic hybridization, are still in their infancy. Two notable examples along this way are the case of Pt ions interacting with Ade-Thy base pairs [122] modifying the hydrogen-bonding architecture, and that reviewed in the previous section of K ions inside the G4 quadruple helix [94]. These metallized DNA structures deserve special attention because they are lately becoming of interest as metallic nanowires [72-74,120-121].

## 3.3 Evaluation of transport through DNA wires based on model Hamiltonians

The available first principle calculations of the electronic properties of DNA molecules, reviewed in the previous section, are complemented by the so-called "model Hamiltonian" studies [123-126]. The latter typically grasp *partial* aspects of the targeted physical system since they are approximate and are not parameter-free theories: parameters are typically fixed by comparison with experiments or with more complex theories such as DFT. On the other hand, model Hamiltonians possess the valuable potential of gaining intuition on the physical mechanisms of the system at hand due to the complexity reduction that they apply. In most cases they provide analytical solutions and allow to control the outcoming physics in the parameter space. Moreover, additional physical effects spanning phonon coupling, electronic correlations, and external driving fields might be added in a modular way.

### 3.3.1 Scattering approach and tight-binding models

The recent progress in nanofabrication unveiled to the experimental investigation the transport properties of structures from the mesoscopic to the molecular scale. Before this was possible, it was already clear that below a certain critical size, the classical behavior of the conductance — scaling proportional to the sample transversal area and inversely proportional to the sample length (Ohm's law) — would break down [127]. The critical size is dominated by the so-called phase-coherence length, that is the length over which quantum coherence is preserved, thus representing a boundary between macroscopic and mesoscopic systems. Moreover, for samples smaller than the electron mean free path (in the so-called ballistic regime) the quantum mechanical wave nature of the electron being transported through the sample turns the classical conduction issue into a quantum wave scattering problem. This idea received its first satisfactory description already in 1957 by Landauer [82] and is also known as "scattering approach". Landauer's theory is by now the privileged tool to describe transport through molecular devices and in particular through dry DNA wires.

The basic components of the scattering approach are the scatterer itself (in our case a sample of molecular size) contacted by two external macroscopic metallic electrodes (which we can conventionally call left and right leads). Figure 20 illustrates a typical two terminal device. The leads are represented by a well defined electrochemical potential, the Fermi level $E_F$, providing a reservoir of electrons at the thermal equilibrium. By slightly biasing the electrochemical potential of one of the two leads with respect to the other, a net current can possibly flow through the sample. At a low bias, the ratio between the current and the bias voltage defines the so-called linear conductance $g$. Following Landauer's intuition, this linear conductance at zero temperature is proportional to the quantum mechanical transmission $T$, by a factor $G_Q = 2e^2/h = (12906\,\Omega)^{-1}$, which is the quantum-of-conductance. This corresponds to writing $g = G_Q T(E_F)$ for electrons injected from a lead with a fixed electrochemical potential $E_F$. The transmission $T(E)$ can be deduced from the scattering matrix (S-matrix) and ultimately from the Hamiltonian of the system after the Fisher-Lee relation [128][1].

---

[1] The key ingredient here is the scatterer Green function, which is the mathematical tool implementing the quantum propagation of an electron through the sample. It is defined as the inverse operator of the scatterer Hamiltonian $H_{sc}$ via the relation $G_{sc}(E) = (E + i0^+ - H_{sc})^{-1}$, and the so-called (left and right) lead self energies $\Sigma_L$ and $\Sigma_R$, which are again obtainable from the Green function of the leads. The Fisher-Lee

It is worth noting that both the Landauer formula and the Fisher-Lee relation are valid in the rather idealized regime in which inelastic scattering through the scattering region can be neglected [83]. Given this idealization, and very close to thermal equilibrium (small voltage drops between the left $E_L$ and the right $E_R$ electrochemical potentials), the current at zero temperature can be estimated as

$$I[V = (E_R/e) - (E_L/e)] = \int_{E_L/e}^{E_R/e} T(E)\,dE .$$

For estimates of the current in cases that are truly out-of-equilibrium, the formalism which we have introduced must be generalized to accommodate phase-correlations (Keldysh formalism) [129]. Such modeling developments are strongly demanded in the near future, to evaluate charge currents through molecular wires under a high bias, given the wide voltage range in the experiments.

Within the Landauer approach, the computation of the quantum conductance is thus traced back to the knowledge of the electronic structure — i.e. the Hamiltonian — of the target system "molecule+leads". The best developed implementations of the Landauer framework employ tight-binding Hamiltonians based on localized orbitals providing the simplest guess for the electronic structure of a molecular system, with parameter optimization from experiments or ab-initio calculations. TB models provide an ideal platform for the computation of the Green function on which the transmission coefficients depend [83]. A generic TB Hamiltonian, to describe a molecular system at hand, is typically written in an orthogonal basis $\{b_{i,\sigma}\}$ that comprises a single maximally compact orbital per atom (taking into account the spin coordinate $\sigma$). In a supramolecular system such as a DNA wire, one orbital per elementary molecule (e.g. the G and C bases in the following example) is included in the basis set. In the TB basis the molecular Hamiltonian is

$$H_{\text{mol}} = \sum_{i,\sigma} \varepsilon_i b_{i\sigma}^\dagger b_{i\sigma} - \sum_{\langle i,j\rangle,\sigma} t_{ij} b_{i\sigma}^\dagger b_{j\sigma} ,$$

---

relation [127] finally reads as $T(E) = 4\,Tr\{\Delta_L(E)\mathcal{G}(E)\Delta_R(E)\mathcal{G}^\dagger(E)\}$. $\mathcal{G}$ is the dressed scatterer Green function (dressed by the presence of the leads) given by $\mathcal{G}^{-1} = \mathcal{G}_{sc}^{-1} + \Sigma_L + \Sigma_R$. $\Delta_\alpha$ ($\alpha = L, R$) is the lead spectral density $\Delta_\alpha = i/2[\Sigma_\alpha(E + i0^+) - \Sigma_\alpha^+(E + i0^+)]$.

where $\varepsilon_i$ are site energies and $t_{ij}$ the hopping matrix elements generally assumed non-zero only between nearest neighbor atomic sites $\langle i, j \rangle$. For an isolated molecule the basis is finite: both the Hamiltonian operator and the associated Green function are represented as matrices. Given the TB Hamiltonian for the molecule and a suitable model for the leads, it is thus possible to estimate the quantum conductance and, to a first approximation, the current-voltage characteristics of the device. The convenient framework provided by TB models for the computation of transport properties will definitely raise more and more interest in the description of transport through single polymers and in particular DNA wires using fast DFT algorithms based on localized orbitals. In this context, it is in fact possible to generate matrix elements that indeed represent a parameter-free version of TB Hamiltonians [130].

### 3.3.2 Applications to poly(dG)-poly(dC) devices

The implementation of the scattering approach and of some simplified electronic structure models for describing the transport behavior of short poly(dG)-poly(dC) DNA wires [14] have been recently independently proposed within two main classes of models. One involves dephasing [124-126] and the other involves the hybridization of the π stack [123].

The Porath et al. experiment [14], reviewed in section 2, reports nonlinear transport measurements on 10.4 nm long poly(dG)-poly(dC) DNA, corresponding to 30 consecutive GC base-pairs, suspended between platinum leads (GC-device). DFT calculations indicated that the poly(dG)-poly(dC) DNA molecule has typical electronic features of a periodic chain [59]. Thus, in both models (assuming dephasing or π–stack hybridization) the poly(dG)-poly(dC) DNA molecule is grained into a spinless linear TB chain. A generalization of the dephasing model to spin-transport has been proposed by Zwolak et al. [124].

In the following we present the key components of these two models and show what consequences their assumptions have on transport.

<u>Dephasing.</u> The idea behind dephasing is that in complex structures, such as DNA, thermal motion and solvation effects might break the phase coherence of the system (a key assumption of the Landauer approach). Within Büttiker's treatment of inelastic scattering [131,132], a possible way to account for dephasing effects in the Landauer approach consists in "inserting" dephasing reservoirs along the scatterer. This has been implemented by Li and Yan [125] by coupling a dephasing term to every site of the spinless linear chain describing the DNA wire. Figure 21 illustrates schematically the diagram for the model system.

Hybridization of the π–stack. In a parallel development, Cuniberti et al. [123] suggested that a plausible mechanism for the observed gap in the Porath et al. experiment [14] could be the hybridization of the overlapping π orbitals in the G-base stack with a perturbation source able to disrupt the metal-like channels. In this case the total Hamiltonian contains one term representing the spinless linear chain (the strong π–stack providing the coupling between any neighboring G-G pairs, as in Figure 22), one term accounting for the hybridization of the π–stack (corrections to the π–stack strong coupling), and one term describing the contacts (Figure 22). The hybridization term, which conserves the phase coherence, physically represents any interaction, internal or external to the molecule itself, which acts to break the perfect structure of the π-stack. Examples of such interactions are: (i) the transversal electronic degrees of freedom at each G base, possibly including the sugar group (Figure 22); (ii) twist effects; (iii) the inherent structure of the double helix which, with electrostatic and Van der Waals interactions, goes beyond a perfect "wirelike" channel where the coupling between neighboring planes may be defined purely within a linear-chain TB description. In the schematic view in Figure 22, all these terms are gathered into the lateral (with respect to the main axis) coupling of the bases with two classes of sites $\alpha = \pm$, but the reader should bear in mind that the microscopic origin of this coupling is not specified at this level and may contain several distinct physico-chemical effects.

Consequences on transport. The dephasing and the π–stack-hybridization are two different mechanisms responsible of the opening of a gap in the transmission. The latter assumes the particularly simple form $T = 4\Delta_L \Delta_R |G_{LR}|^2$ due to the low-complexity of the underlying Hamiltonians [83]. Here, $\Delta_\nu = -\text{Im}(\Sigma_\nu)$ is the spectral density of the metal-molecule coupling at the left (L) and right (R) leads.

$G_{LR}$ is the molecular matrix element of the Green function between the two contact sites dressed by the lead self-energy[2].

Starting from the knowledge of the transmission $T(E)$, the calculation of the current can be pursued within the scattering formalism as presented in the previous section, and the results for the two models show an overall agreement with the experimental findings (see Fig. 23).

Despite the different assumptions in the two schemes described here, the properties of the calculated currents share the expected semiconducting features. This is basically due to the fact that they are formally acting on the π−stack in a similar way. Only further experimental investigation could shed more light on the discrimination of the gap-opening mechanism. Such an effort would eventually help in quantifying the influence of other insulating effects, which are possibly present in other DNA-based devices. They include (i) electron correlations and Coulomb blockade, (ii) localization effects due to the sequence variability, and (iii) local defects.

Although based on a simplified parametric description of the electronic structure of the molecule and of the leads, the framework discussed in this section has the advantage of leading directly to the computation of measurable quantities (the *I-V* curves). Thus, it is possible to relate the experimental observations to the quantum-mechanical properties of the systems under investigation, e.g. the electronic energy-level-structure of the molecule and the relation of such levels to the energy of the leads. A timely improvement in this direction will come from the implementation of manageable methods, which combine a parameter-free atomistic description of the electronic structure of the molecular device [85-88,130] with the well established Green function formalism to compute the transport characteristics. In such a way, it would be possible to distinguish self-consistently in the calculations different mechanisms, and their relative importance in experimental settings.

---

[2] Interestingly enough the transmission calculation can be pursued analytically. In the particular case of the π−stack hybridized model [122], the gap in the transmission is a function only of the longitudinal and transverse hopping, $\Delta_T = 2\sqrt{t_\parallel^2 + 2t_\perp^2} - 2t_\parallel$. Note that in the case of non-hybridized π−stack $t_\perp = 0$ the gap is closed, as we could expect. The gap opening mechanism (common feature of the two models) can also be understood by referring to their dispersion relations also with a gap at the Fermi level.

## 4 Conclusions and Perspectives

Charge migration along DNA molecules has attracted a considerable scientific interest for over half a century. Results of solution chemistry experiments on large numbers of short DNA molecules indicated high charge-transfer rates between a donor and an acceptor located at distant molecular sites. This, together with the extraordinary molecular recognition properties of the double helix and the hope of realizing the bottom-up assembly of molecular electronic devices and circuits using DNA molecules, have triggered a series of direct electrical transport measurements through DNA. In this chapter we provided a comprehensive review of these measurements and of the intense theoretical effort, in parallel and following the experiments, to explain the results and predict the electronic properties of the molecules, using a variety of theoretical methods and computational techniques.

After the appearance of some initial controversial reports on the conductance of DNA devices, recent results seem to indicate that native DNA does not possess the electronic features desirable for a good molecular electronic building block, although it can still serve as a template for other conducting materials [12,13,133]. Particularly, it is found that short DNA molecules are capable of transporting charge carriers, and so are bundles and DNA-based networks. However, electrical transport through long single DNA molecules that are attached to surfaces is blocked, possibly due to this attachment to the surface.

In a molecular electronics perspective, however, the fascinating program of further using the "smart" self-assembly capabilities of the DNA to realize complex electronic architectures at the molecular scale remains open for further investigation. To enhance the conductive properties of DNA-based devices there have been already interesting proposals for structural manipulation. These include intrinsic doping by metal-ions incorporated into the double helix (such as M-DNA [72-75,120-121]), exotic structures like G4-DNA [94,119,134], along with the synthesis of other novel helical structures. The reach zoology of DNA derivatives with more promising electrical performance will definitely represent a great challenge in the agenda of researchers involved with the transport in DNA wires.

A last comment is due on needs for progress in the investigation tools. Further development in the study of potential DNA nanowires requires the advancement of synthesis procedures for the structural modifications, and an extensive effort in X-ray and NMR characterization. Concerning direct conductivity measurements, techniques for deposition of the molecules onto inorganic substrates and between electrodes must be optimized. Moreover, the experimental settings and contacts must be controlled to a high degree of accuracy in order to attain an uncontroversial interpretation and high reproducibility of the data. On the theoretical side, a significant breakthrough might be the combination of mesoscopic theories for the study of quantum conductance and first-principle electronic structure calculations, suitable for applications to the complex molecules and device configurations of interest. Given this background, we believe that there is still plenty of room to shed light onto the appealing issue of charge mobility in DNA, for both the scientific interest in conduction through one-dimensional polymers and the nanotechnological applications. The high interdisciplinary content of such a research manifesto will necessarily imply a crossing of the traditional borders separating solid-state physics, chemistry and biological physics.

## Acknowledgements

Funding by the EU through grant FET-IST-2001-38951 is acknowledged. DP is thankful to Cees Dekker and his group, with whom experiment [14] was done, to Joshua Jortner, Avraham Nitzan, Julio Gomez-Herrrero, Christian Schönenberger and Hezy Cohen for fruitful discussions about the conductivity in DNA and critical reading of the manuscript. DP research is funded by: The First foundation, The Israel Science Foundation, The German-Israel Foundation and Hebrew University Grants. GC acknowledges the collaboration with Luis Craco with whom part of the work reviewed was done. The critical reading of Miriam del Valle, Rafael Gutierrez, and Juyeon Yi is also gratefully acknowledged. GC research has been funded by the Volkswagen Foundation. RDF is extremely grateful to Arrigo Calzolari, Anna Garbesi, and Elisa Molinari, for fruitful collaborations and discussions on topics related to this chapter, and for a critical reading of the manuscript. RDF research is funded by INFM through PRA-

# Figure Captions

**Figure 1.** (a-c) 16 μm long λ-DNA was stretched between two metal electrodes using short hang-over single strands complementary to single-strands that were pre-attached to the metal electrodes. (d) A fluorescent image of the DNA molecule, connecting the metal electrodes. (e) The flat, insulating current-voltage that was measured. (from [12], by permission; © 1998 by Nature Macmillan Publishers Ltd).

**Figure 2.** (a) The LEEPS microscope used to investigate the conductivity of DNA. The atomic-size electron point source is placed close to a sample holder with holes spanned by DNA molecules. Due to the sharpness of the source and its closeness to the sample, a small voltage $U_e$ (20-300 mV) is sufficient to create a spherical low-energy electron wave. The projection image created by the low-energy electrons is observed at a distant detector. Between the sample holder and the detector, a manipulation-tip is incorporated. This tip is placed at an electrical potential $U_m$ with respect to the grounded sample holder and is used to mechanically and electrically manipulate the DNA ropes that are stretched over the holes in the sample holder. (b) A projection image of λ-DNA ropes spanning a 2-μm-diameter hole. The kinetic energy of the imaging electrons is 70 eV. (c) SEM image, showing the sample support with its 2-μm-diameter holes. (d) SEM image of the end of a tungsten manipulation-tip used to contact the DNA ropes. Scale bar 200 nm. (e) The metal tip is attached to the λ-DNA molecule. (f) *I-V* curves taken for a 600-nm-long DNA rope. In the range of ± 20 mV, the curves are linear; above this voltage, large fluctuations are apparent. A resistance of about 2.5 MΩ was derived from the linear dependence at low voltage (from [50], by permission; © 1999 by Nature Macmillan Publishers Ltd).

**Figure 3.** (a) Current-voltage curves measured at room temperature on a 10.4-nm-long DNA molecule [30 base-pairs, double-stranded poly(dG)-poly(dC)] trapped between two metal nanoelectrodes that are 8 nm apart. Subsequent *I-V* curves (different colors) show similar behavior but with a variation of the width of the gap. The upper inset shows a schematic of the sample layout. Using electron-beam lithography, a local 30 nm narrow segment in a slit in the SiN layer is created. Underetching the $SiO_2$ layer leads to two opposite freestanding SiN "fingers" that become the metallic nanoelectrodes after sputtering Pt through a Si mask. The lower inset is a SEM image of the two metal electrodes (light area) and the 8 nm gap between them (dark area). Deposition of a DNA molecule between the electrodes was achieved with electrostatic trapping. A 1 μl droplet of dilute DNA solution is positioned on top of the sample. Subsequently, a voltage of up to 5 V is applied between the electrodes. The electrostatic field polarizes a nearby molecule, which is then attracted to the gap between the electrodes due to the field gradient. When a DNA molecule is trapped and current starts to flow through it, a large part of the voltage drops across a large (2 GΩ)

series resistor, which reduces the field between the electrodes and prevents other molecules from being trapped. Trapping of DNA molecules using this method is almost always successful. (b) Current-voltage curves that demonstrate that transport is indeed measured on DNA trapped between the electrodes. The solid curve is measured after trapping a DNA molecule as in (a). The dashed curve is measured after incubation of the same sample for 1 hour in a solution with 10 mg/ml DNase I enzyme. The clear suppression of the current indicates that the double-stranded DNA was cut by the enzyme. This experiment was carried out for 4 different samples (including the sample of Fig. 4). The inset shows two curves measured in a complementary experiment where the above experiment was repeated but in the absence of the Mg ions that activate the enzyme and in the presence of 10 mM EDTA (ethylenediamine tetraacetic acid) that complexes any residual Mg ions. In this case, the shape of the curve did not change. This observation verifies that the DNA was indeed cut by the enzyme in the original control experiment (from [14], by permission; © 2000 by Nature Macmillan Publishers Ltd).

**Figure 4.** Differential conductance $dI/dV$ versus applied voltage V at 100 K. The differential conductance manifests a clear peak structure. Good reproducibility can be seen from the six nearly overlapping curves. Peak structures were observed in four samples measured at low temperatures although details were different from sample to sample. Subsequent sets of *I-V* measurements can show a sudden change, possibly due to conformational changes of the DNA. The inset shows an example of two typical *I-V* curves that were measured before and after such an abrupt change. Switching between stable and reproducible shapes can occur upon an abrupt switch of the voltage or by high current (from [14], by permission; © 2000 by Nature Macmillan Publishers Ltd).

**Figure 5.** AFM images of DNA assembled in various devices. (a) Mixed-sequence DNA between platinum electrodes spaced by 40 nm. Scale bar 50 nm. (b) Height image of poly(dG)-poly(dC) DNA bundles on platinum electrodes. The distance between electrodes is 200 nm, and the scale bar is 1 μm. (c) High magnification image of the device shown in b. Several DNA bundles clearly extend over the two electrodes. Scale bar 200 nm. (d) Poly(dG)-poly(dC) DNA bundles on platinum electrodes fabricated on a mica substrate. Scale bar 500 nm. For all these devices, no conduction was observed (from [60], by permission; © 2001 by the American Institute of Physics).

**Figure 6.** (a) Schematic drawing of the measured sample, with DNA molecules combed between Re/C electrodes on a mica substrate. (b) AFM image showing DNA molecules combed on the Re/C bilayer. The large vertical arrow indicates the direction of the solution flow. The small arrows point towards the combed molecules. Note the forest structure of the carbon film. (c) DC resistance as a function of temperature on a large temperature scale for three different samples, showing the power law behavior down to 1 K (from [61], by permission; © 2001 by Science).

**Figure 7.** (a) Three-dimensional SFM image showing two DNA molecules in contact with the gold electrode. The image size is 1.2 μm$^2$. A scheme of the electrical circuit used to measure the DNA resistivity is also shown. (b) λ-DNA strands connecting two gold electrodes spanned on a bare mica gap. The image analysis leads to the conclusion that at least 1000 DNA molecules are connecting the electrodes. From the (absence of) current between the electrodes, a lower bound of 10$^5$ Ωcm per molecule is obtained for the resistivity of DNA at a bias voltage of 10 V (from [59], by permission; © 2000 by the American Physical Society).

**Figure 8.** (a) Schematic of the electric current measurement. Two CNT probes (p1 and p2) of the nanotweezers were set on a DNA. In a two-probe dc measurement, one of the CNT probes (p1) was used as the cathode. A CNT-AFM probe was contacted with the DNA as the anode. The electric current was measured while varying the distance between the anode and cathode $d_{CA}$. (b) AFM image (scale bar, 10 nm) of a single DNA molecule attached with two CNT probes (p1 and p2) of the nanotweezers, which was obtained by scanning the CNT-AFM probe. (c) $d_{CA}$-dependence of electric current ($I_{CA}$) between the electrodes measured with the electrode configuration shown in (a). (d) Schematic of the electric measurement under applied gate bias. Two nanotweezer probes (p1 and p2) and the CNT-AFM probe were used as source, drain, and gate electrodes, respectively. The electric current between the source and drain was measured with varying the distance between source and gate $d_{GS}$. (e) $d_{GS}$-dependence of the electric current ($I_{DS}$) between the source and the drain electrodes measured with the electrode configuration shown in (d). The solid gray line shows the electric current for the CNT-AFM probe moving from the source electrode to the drain electrode. Dashed black line is for the case of the opposite moving direction (from [63], by permission; © 2003 by the American Institute of Physics).

**Figure 9.** (a) Schematic illustration of the measurement with a conducting-probe AFM. (b) Relationship between resistance and DNA length. The exponential fitting plots of the data are also shown. (c) Typical *I-V* curves of poly(dG)-poly(dC), the linear Ohmic behaviors on L=100 nm at the repeat measurement of five samples. (d) Rectifying curves of poly(dG)-poly(dC) at L=100 nm (from [68], by permission; © 2000 by the American Institute of Physics).

**Figure 10.** (a) Current-voltage curves measured in vacuum at room temperature on M-DNA (○) and B-DNA (●) molecules. The DNA fibers are 15 μm long and the interelectrode spacing is 10 μm. In contrast to the B-DNA behavior, M-DNA exhibits no plateau in the *I-V* curve. The lower inset shows the schematic experimental layout. The upper inset shows two representative current-voltage curves measured in vacuum at room temperature on samples of Au–oligomer–B-DNA–oligomer–Au in series. (b) AFM Image of a M-DNA bundle on the surface of the gold electrode (scale bar: 1 μm). (c) Cross section made along the white line in (b) using tapping-mode AFM giving a bundle height of 20–30 nm and width of about 100 nm, which implies it consists of ~300 DNA strands (from [72], by permission; © 2001 by the American Physical Society).

**Figure 11.** (a) SEM image of an Au/Ti nanoelectrode with a 20 nm spacing. Three electrodes are shown, S and D stand for source and drain. (b) *I-V* curves measured at room temperature for various values of the gate voltage ($V_{gate}$) for poly(dG)-poly(dC). The inset of (b) is the schematic diagram of electrode arrangement for gate dependent transport experiments. (c) Conductance versus inverse temperature for poly(dA)-poly(dT) and poly(dG)-poly(dC), where the conductance at *V*=0 was numerically calculated from the *I-V* curve (from [76], by permission; © 2001 by the American Physical Society).

**Figure 12. a.** Schematic illustration of a flexible, aligned DNA film prepared from casting organic-soluble DNA-lipid complexes with subsequent uniaxial stretching. **b.** Experimental geometries and measured dark currents for aligned DNA films (20x10 mm, thickness 30±5 μm) on comb-type electrodes at 25 °C. In the dark-current plot, the three curves represent different experimental settings and environments: (a) DNA strands in the film placed perpendicular to the two electrodes (scheme in the upper inset) and measured in ambient; (b) the same film as in (a) measured in a vacuum at 0.1 mmHg; (c) DNA strands in the film placed parallel to the two electrodes, both in a vacuum and in ambient (from [49], by permission; © 1998 by the American Chemical Society).

**Figure 13.** (a) Schematic of the device used in the experiment. (b) SEM image of the gold nanoelectrodes, fabricated by electron beam lithography and lift-off onto a $SiO_2$/Si substrate. (c) Schematics of the ribbon-like structure formed by the deoxyguanosine molecules connected through hydrogen-bonds. R is a radical containing the sugar and alkyl chains. (d) AFM micrograph of the ordered deoxyguanosine film obtained after drying the solution in the gap. Regular arrangement of ribbons ranges over a distance of about 100 nm. The ribbon width of about 3 nm is consistent with that determined by X-ray measurements. (e) *I–V* characteristics of the device (from [69], by permission; © 2001 by the American Institute of Physics).

**Figure 14.** (a) Isosurface plot of the HOMO of an isolated G molecule, exhibiting a clear π character. (b) Pictorial illustration of the construction of a model periodic wire through the stacking of G bases. To model the periodic structures, Di Felice et al. [91] employed a unit supercell containing two G molecules, including only nearest-neighbor interactions. The models do not reproduce the continuous helicity characteristic of a G strand in DNA. The helicity along the axis of the pile goes instead back and forth at successive steps: if the rotation between planes N-1 and N is Θ, that between planes N and N+1 is - Θ, and plane N+1 is equivalent to N-1. Since nearest-neighbor coupling is already very weak and configuration-dependent, it is expected that longer-range interactions would not contribute effectively to orbital delocalization.

**Figure 15.** Bandstructure of periodic model G-wires with different angles Θ (defined in Figure 14), computed between Γ and the A edge of the 1D BZ. The bandstructure for each system was calculated at several **k** points of the 1D BZ along the stack axis, interpolating from the self-consistent charge density obtained at one special **k** point. The insets show the top-view geometries of the selected periodic stacks: gray and black spheres indicate atoms on two consecutive planes. Left: Θ=0º, the ΓA dispersion of the HOMO-derived (LUMO-derived) band is 0.65 eV (0.52 eV upwards). Middle: Θ=180º, the ΓA dispersion of the HOMO-derived (LUMO-derived) band is 0.26 eV downwards (0.13 eV upwards). Right: Θ=36º, the ΓA dispersion for the HOMO- and LUMO-derived bands is vanishing, index of poor interaction. For Θ=36º the π overlap between consecutive stacked planes is negligible and the energy levels remain flat and degenerate. The π overlap becomes larger for Θ=180º and is maximum for Θ=0º: this is reflected in an increasing band dispersion for both the valence and conduction bands. The effective masses obtained by a parabolic interpolation of the top valence and bottom conduction bands are in the range of 1-2 free electron masses for Θ=0º and Θ=180º (typical of wide-band-gap semiconductors). The DFT energy gap (underestimated [97]) is in the range of 3.0÷3.5 eV (adapted from [91], by permission; © 2002 by the American Physical Society).

**Figure 16.** Inner-core structure of the computed G4-wires. (a) A palanar tetrad G4, which constitutes the basic building block for stacking planes. (b) Pictorial illustration of the rotation between adjacent stacked planes: each plane is represented as a square and each step along the quadruple helix the plane is rotated by 90º, corresponding to a 30º rotation between individual G bases. (c) Atomic model of the periodicity unit in the computed G4-wires. (d) Pictorial illustration of the G4-helix molecules in the crystal structure. The crystallized quadruplexes [119] are finite molecules containing eight G4 planes, separated in two sets of four G4 planes : within one set (e.g. upper four planes in panel d), consecutive planes are connected via a sugar-phosphate backbone similar to that in DNA, whereas the backbone is interrupted and the polarity is inverted at the interface between two different sets (see the break between the upper and lower four planes in panel d). In the computational work by Calzolari et al. [94], a periodic infinite (1D crystal-like) wire was constructed from three (see panel c) of the four planes in one of the two sets, based on the evidence that the fourth plane in the stack is equivalent to the first one by symmetry. The external backbone was neglected in the simulations. In the central cavity of the tubular sequence, one $K^+$ ion was inserted between any two planes, as suggested from the X-ray analysis ($Na^+$ ions were present in the crystals) (adapted from [94], by permission; © 2002 by the American Institute of Physics).

**Figure 17.** Bandstructure (left) and DOS (right) computed for the G4-wire. In the DOS, the shaded (white) region pertains to occupied (empty) electron states. The labels π and σ identify the orbital character. Although the bandstructure obtained with translational symmetry identifies the localized character of the one-particle electron wavefunctions, the calculated DOS is consistent with a model of a semiconducting nanowire, which hosts extended channels for electron/hole

motion. The peaks in the effective band-like DOS result from the energy spreading of the single-molecule energy levels that group into manifolds (see the groups of flat bands in the left panel) (adapted from [94], by permission; © 2002 by the American Institute of Physics).

**Figure 18.** Contour (left) and isosurface (right) plots of the HOMO-band convolution of electron states for the G4-DNA base-core structure [94]. The delocalized character along the guanines is evident, as highlighted by the arrow parallel to the axis of the stack (adapted from [94], by permission; © 2002 by the American Institute of Physics).

**Figure 19.** Schematic level diagram around the Fermi level, for the infinite DNA wire studied by Gervasio et al. [95]. The Fermi level positioned in the middle of the gap was chosen as the zero of energy. The highest occupied "band" is constituted of a manifold of 12 states localized on the 12 Gua bases contained in the periodic unit and originated from the Gua-HOMO. Cyt-localized states ($\pi^*$ Cyt) appear as another manifold at 3.16 eV above the HOMO-band. However, empty electron states due to metal counterions and phosphates are revealed at 1.28 eV above the HOMO-band (see also [93]), and the ground-to-excited-state transitions are therefore related to charge transfer between the inner and outer helix (adapted from [95] by permission; © 2002 by the American Physical Society).

**Figure 20.** Schematic representation of a two-terminal device. The scattering region (enclosed in the dashed-line frame) with transmission probability $T(E)$ is connected to semi-infinite left (L) and right (R) leads which end into electronic reservoirs (not shown) at chemical potentials $E_L$, and $E_R$, kept fixed at the same value $E_F$ for linear transport. By applying a small potential difference electronic transport will occur. The scattering region or molecule may include in general parts of the leads (shaded areas) (adapted from [106] by permission; © 2002 by Springer Verlag).

**Figure 21.** The DNA molecule described by a one-dimensional TB Hamiltonian consists of a stack of GC pairs (the circles). To simulate the phase-breaking effect, each GC pair is connected with a dephasing reservoir (the oval) with an Hamiltonian $H_{deph} = \sum_i \Sigma_i b_i^\dagger b_i$. Following Ref. [132], all dephasing self-energies $\Sigma_i$ have a non-smooth dependence on energy: $\Sigma_i = \eta^2 / [E - \varepsilon_i - \sigma_i(E)]$, where $\sigma_i = (E - \varepsilon_i)/2 - i[t^2 - (E - \varepsilon_i)^2 / 2]^{1/2}$; the quantities $\varepsilon_i$ and $t$ are the linear chain TB parameters, and $\eta$ is a parameter controlling the dephasing intensity. The total Hamiltonian thus reads as $H = H_{mol} + H_{deph} + H_{cont}$ where the contact Hamiltonian $H_{cont}$ accounts for the leads and of their coupling to the molecule in a standard way [106] (from [125] by permission; © 2002 by the American Institute of Physics).

**Figure 22.** Schematic view (left) of a fragment of poly(dG)-poly(dC) DNA molecule; each GC base pair is attached to sugar and phosphate groups forming the molecule backbone. On the right side, the diagram of the lattice adopted in building our model, with the π stack connected to the isolated states denoted as ±-edges. The total Hamiltonian reads $H = \sum_i \varepsilon_i b_i^\dagger b_i - \sum_{\langle i,j \rangle} t_\parallel b_i^\dagger b_j + \sum_{i,\alpha} \varepsilon_\alpha c_{i\alpha}^\dagger c_{i\alpha} - t_\perp \sum_{i,\alpha=\pm} \left( c_{i\alpha}^\dagger b_{i\sigma} + h.c. \right) + H_{cont}$, where the first two terms represent the spinless linear chain (the strong π–stack providing the coupling between any neighboring G-G pairs) and the second two accounts for the hybridization of the π–stack with the disconneted bands represented by the operator $c_{i\alpha}$. $H_{cont}$ is the metal-molecule coupling Hamiltonian (from [123] by permission; © 2002 by the American Physical Society).

**Figure 23.** (a) *I-V* characteristics from the work by Lee and Yang [125]: theoretical results versus experimental data. The transport currents in the presence of weak dephasing ($\eta = 0.05$ eV) and a stronger one ($\eta = 0.3$ eV) are theoretically shown by the solid and dashesd curves. (b) Low temperature *I-V* characteristics of two typical measurements for a 30-base-pair poly(dG)-poly(dC) molecule at 18 K (blue circles) and at 3.6 K (red circles) [123]. Solid lines show the theory curves following the experimental data. The insets show the transmission calculated after the blue data (upper) and the normalized differential conductance (lower). The parameters used (see Figure 21) are $t_\parallel = 0.37$ eV and $t_\perp = 0.74$ eV for the blue measurement, and $t_\parallel = 0.15$ eV and $t_\perp = 0.24$ for the red one. These values are considered for a homogeneous system. If one draws the transversal coupling parameters for a disordered system their average value would be lower. For a related experiment supporting this idea cf. [114] (from [125] by permission; © 2001 by the American Institute of Physics; from [123], by permission; © 2002 by the American Physical Society).

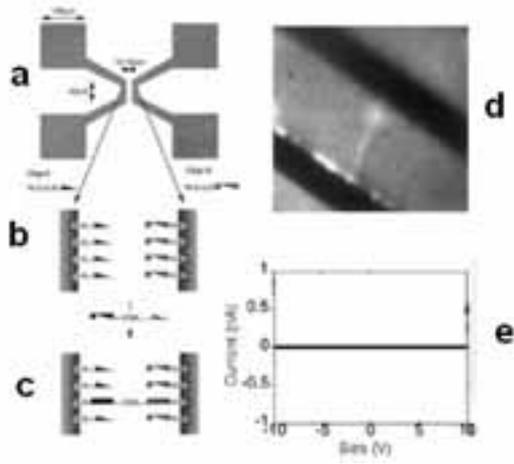

**Figure 1**

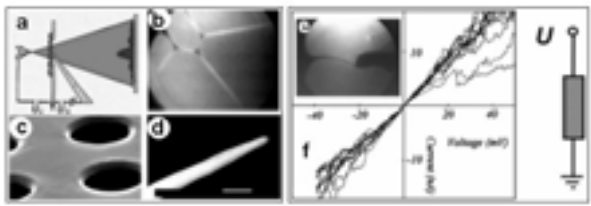

**Figure 2**

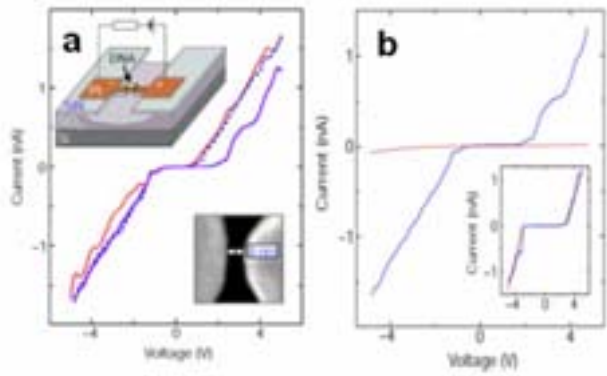

**Figure 3**

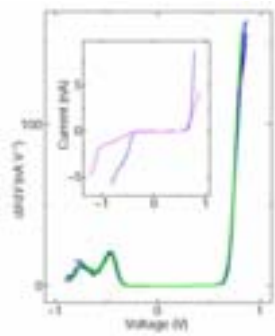

**Figure 4**

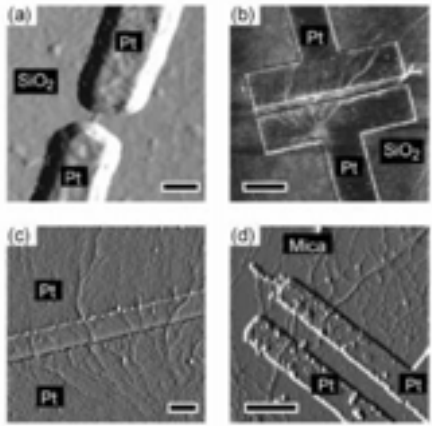

**Figure 5**

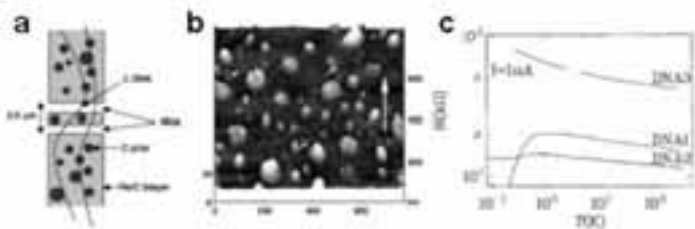

**Figure 6**

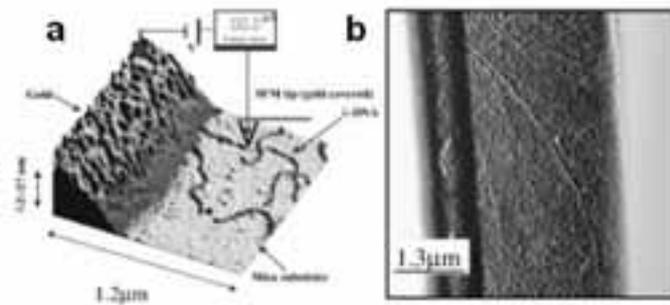

**Figure 7**

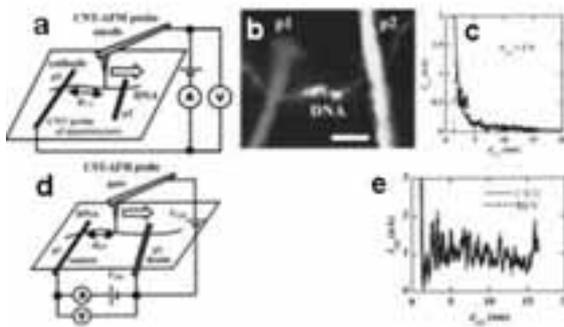

**Figure 8**

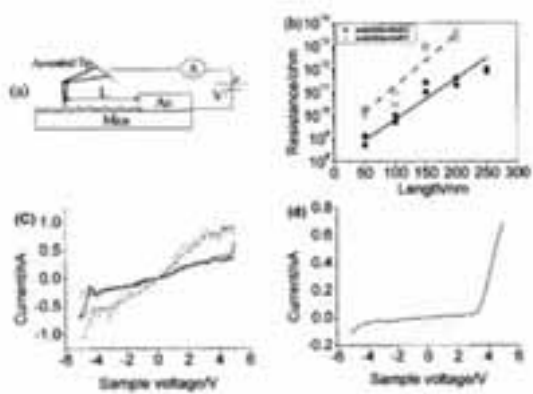

**Figure 9**

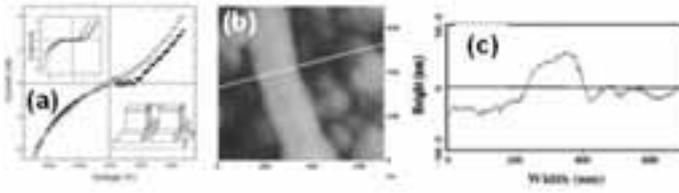

**Figure 10**

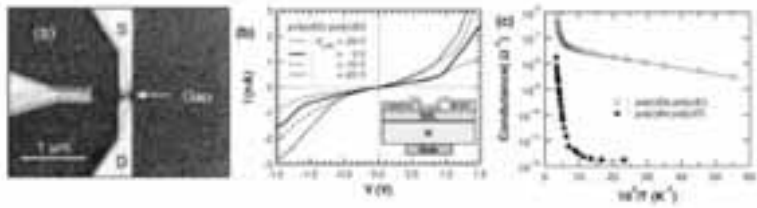

**Figure 11**

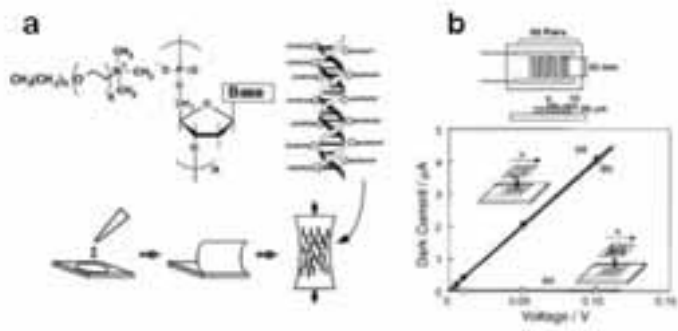

**Figure 12**

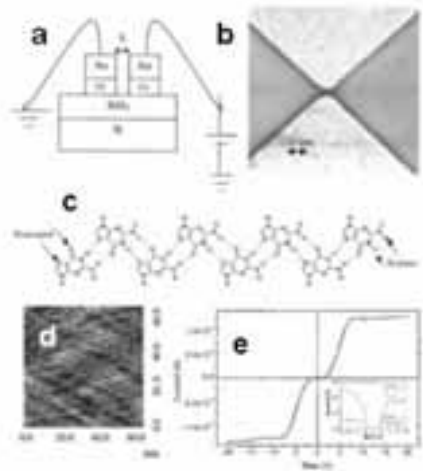

**Figure 13**

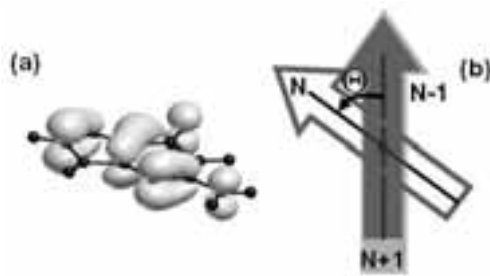

**Figure 14**

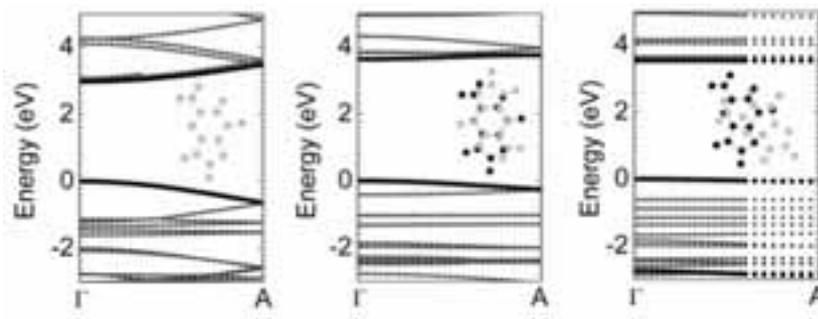
**Figure 15**

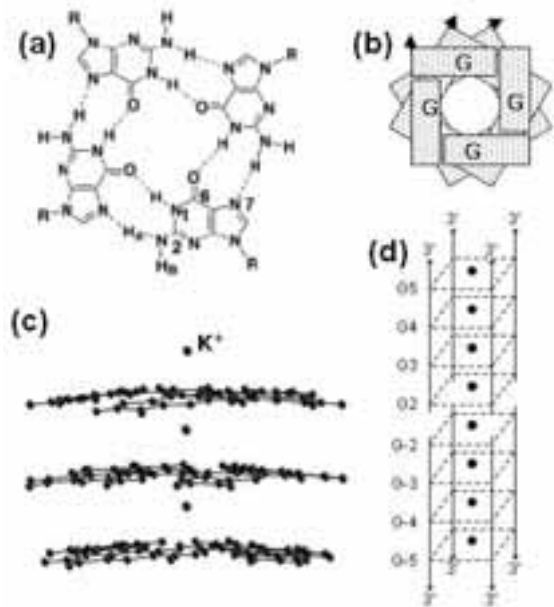
**Figure 16**

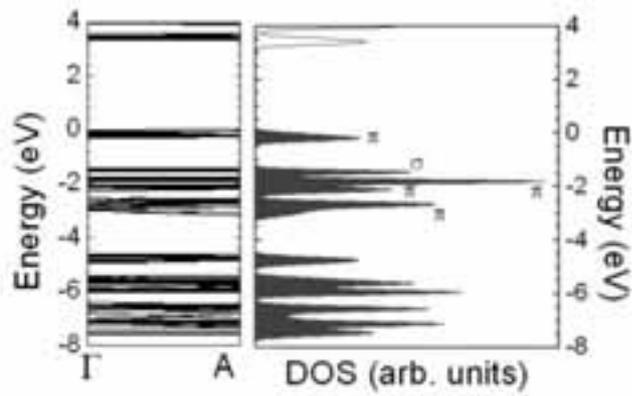
**Figure 17**

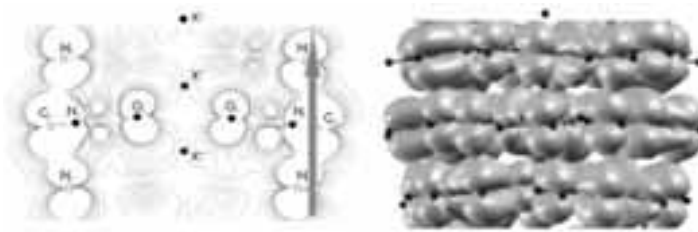
**Figure 18**

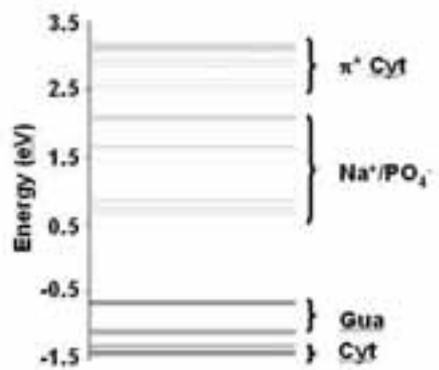

**Figure 19**

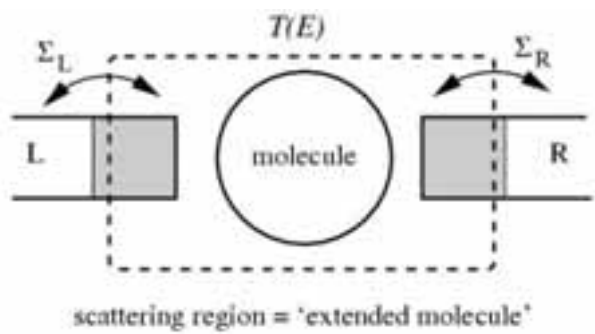

**Figure 20**

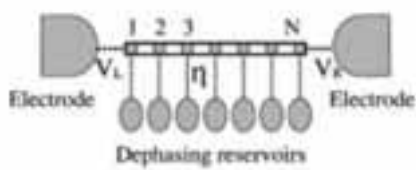

**Figure 21**

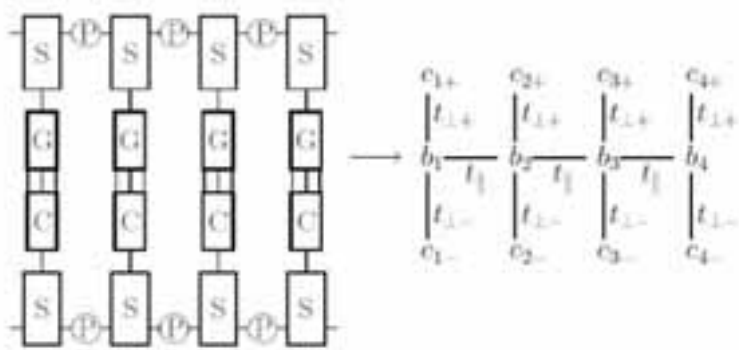

**Figure 22**

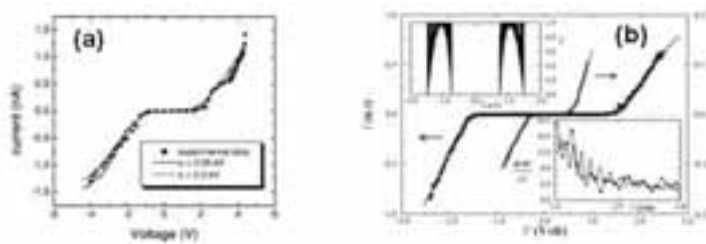

**Figure 23**